\documentclass[]{spie}  

\usepackage{amsmath,amsfonts,amssymb}
\usepackage{graphicx}
\usepackage{subfigure}
\usepackage{float}
\usepackage{todonotes}
\usepackage[colorlinks=true, allcolors=blue]{hyperref}
\usepackage[normalem]{ulem}
 
\title{The Canadian Galactic Emission Mapper (CGEM): A Cosmic Microwave Foreground Experiment}

\author[a]{Parham Zarei}
\author[a]{Pedro Villalba-González}
\author[a]{Mandana Amiri}
\author[b]{Charles L. Bennett}
\author[a]{Guinevere Berg}
\author[a]{Mark Halpern}
\author[a]{Gary Hinshaw}
\author[c]{Gordon Lacy}
\author[a,c]{Joshua MacEachern}
\author[a]{Thomas J. Rennie}
\author[a]{Shuyu van Kerkwijk}
\author[c]{Bruce Veidt}
\author[b]{Janet Weiland}
\author[a]{Don Wiebe}
\author[d]{Edward J. Wollack}

\affil[a]{Department of Physics and Astronomy, University of British Columbia, Vancouver, BC V6T 1Z1, Canada}
\affil[b]{The William H. Miller III Department of Physics and Astronomy,
Johns Hopkins University, 3400 N. Charles Street, Baltimore, MD 21218 USA}
\affil[c]{Dominion Radio Astrophysical Observatory, Herzberg Astronomy \& Astrophysics Research Centre, National Research Council Canada,
Penticton, BC, Canada}
\affil[d]{NASA Goddard Space Flight Center, 8800 Greenbelt Road, Greenbelt, MD 20771, USA}
\authorinfo{Further author information: (Send correspondence to P.Z or P.V.G.)\\P.Z: E-mail: pzarei@phas.ubc.ca\\  P.V.G.: E-mail: pedrovg@phas.ubc.ca}

\begin{document} 
\maketitle

\begin{abstract}
The Canadian Galactic Emission Mapper (CGEM) is a 4-meter single-dish radio telescope located at the Dominion Radio Astrophysical Observatory (DRAO) in Penticton, Canada. CGEM is designed to map polarized Galactic synchrotron emission across the entire northern sky at 8-10 GHz with 1 MHz frequency resolution and $\sim0.5$ degree angular resolution. Its goal is to obtain high-fidelity, low-noise maps of polarized Galactic synchrotron radiation, at frequencies where synchrotron dominates, in order to provide a reliable template for CMB $B-$mode foreground cleaning. We will show on sky performance of CGEM’s single-pixel azimuthally-symmetric telescope, its coherent, dual-polarized radiometer, and show maps made from commissioning data from the first few months of operation.
\end{abstract}



\section{INTRODUCTION}
\label{sec:intro}  

\begin{figure}[t!]
 		\includegraphics[width=\linewidth]{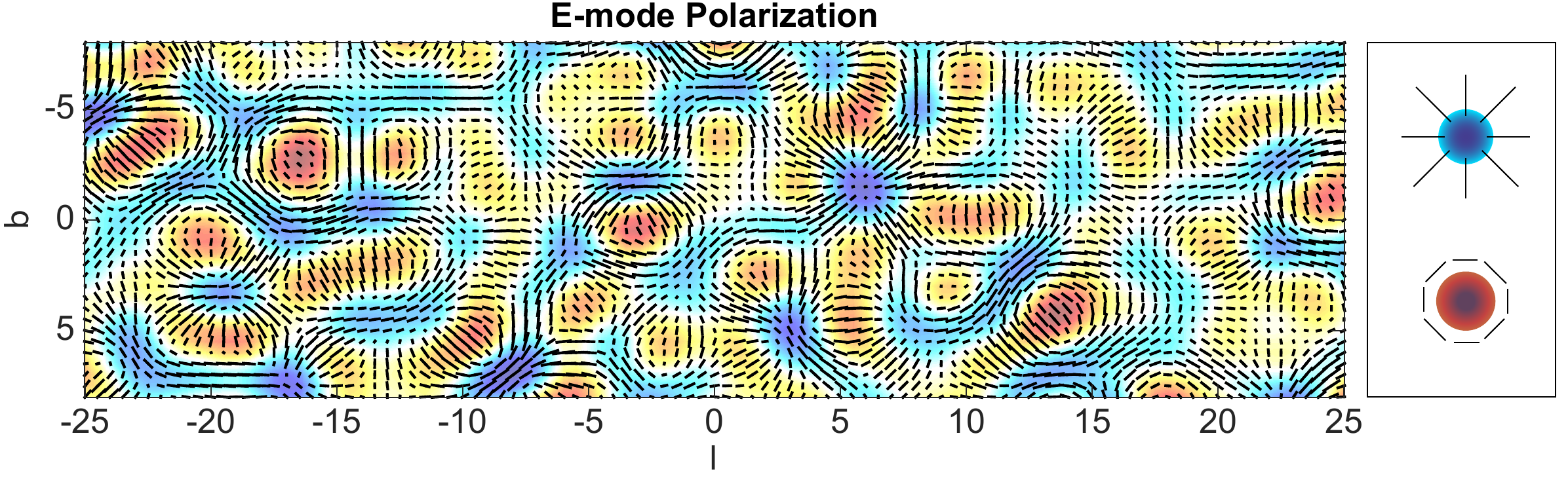}
 		\includegraphics[width=\linewidth]{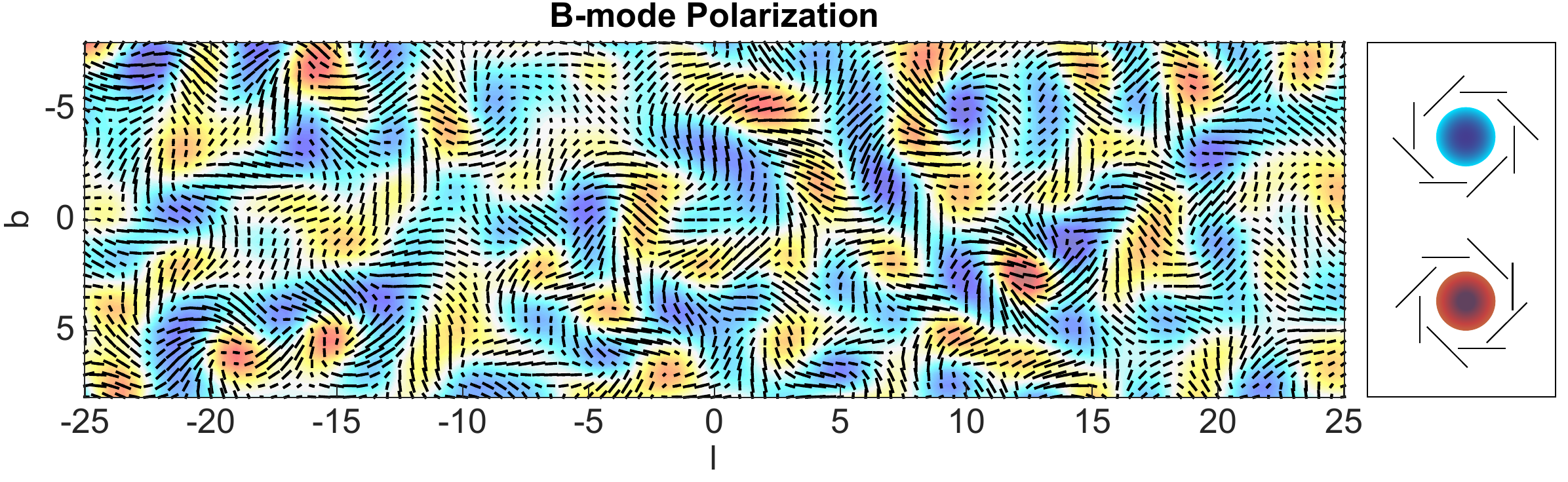}
 		\caption{\textbf{$E$- and $B$-mode polarization pattern sketch.} Polarization patterns are shown as lines superimposed on a colour map of intensity anisotropy of the CMB. The $E$-mode polarization pattern is tangential in hot spots (red) and radial around cold spots (blue). In the case of $B$-modes there is a swirling pattern around hot and cold spots. The image shows them in the same scale for illustration purposes, but anticipated $B-$modes are at least two orders of magnitude fainter than $E-$modes. (Image from~\citenum{kamionkowski2016quest})}
 		\label{fig:pureGC}
 	\end{figure}

Cosmic inflation predicts that the universe underwent a period of exponential expansion in the first fraction of a second after the Big Bang \cite{guth1981inflationary,linde1982new}. This inflationary epoch would solve several fundamental problems in cosmology, such as the horizon, flatness, and monopole problems \cite{dodelson2020modern,weinberg2008cosmology}. Had inflation occurred, it is theorized that primordial gravitational waves sourced at inflation would have left an imprint on the polarization field of the Cosmic Microwave Background (CMB) in the form of $B-$mode polarization patterns \cite{baumann2022cosmology}. The polarization field of the CMB can be decomposed into a curl-free ($E-$) and a divergence-free ($B-$) mode through the Helmholtz decomposition \cite{zaldarriaga1997all,kamionkowski1997statistics}. While scalar perturbations in the early universe only source $E-$modes, tensor perturbations (which propagate as gravitational waves) source both $E-$ and $B-$modes. See figure~\ref{fig:pureGC} for a visualization of each mode and representative patterns on the sky. The detection of these primordial $B-$modes would provide the strongest evidence to date for inflation and would probe physics at energy scales near $10^{16}$ GeV ---far beyond the reach of any experiment we might conduct on Earth \cite{kamionkowski2016quest}. The $E$-mode signal has been measured and characterized by multiple experiments, see for instance~\citenum{Dutcher2021}. Unfortunately, after decades of observational efforts, $B-$modes are not detected; an upper limit in the ratio of $E-$ to $B-$ mode power has been estimated to be $r\sim0.036$, proving how challenging the hunt for the cosmic $B-$mode signal is \cite{Ade_2021}.
Aside from the predicted low intensity of this signal, there are several challenges associated with the search for $B-$modes. First and foremost, many sources between an observer and the CMB (hereafter foregrounds) are also polarized and confuse our view of the CMB polarization. These foregrounds, and in particular our Galaxy, are much brighter than the expected $B-$mode signal, and thus need to be mapped with extremely high precision and subsequently removed from CMB data. The false detection of cosmological $B-$modes by the BICEP2 collaboration in 2014, later attributed to polarized dust foreground contamination, underscores the critical importance of accurate foreground characterization  \cite{ade2014detection,flauger2014toward}.

\section{CMB Galactic Foregrounds}
\begin{figure}[t!]
	\centering
	\subfigure{
		\includegraphics[width=0.49\textwidth]{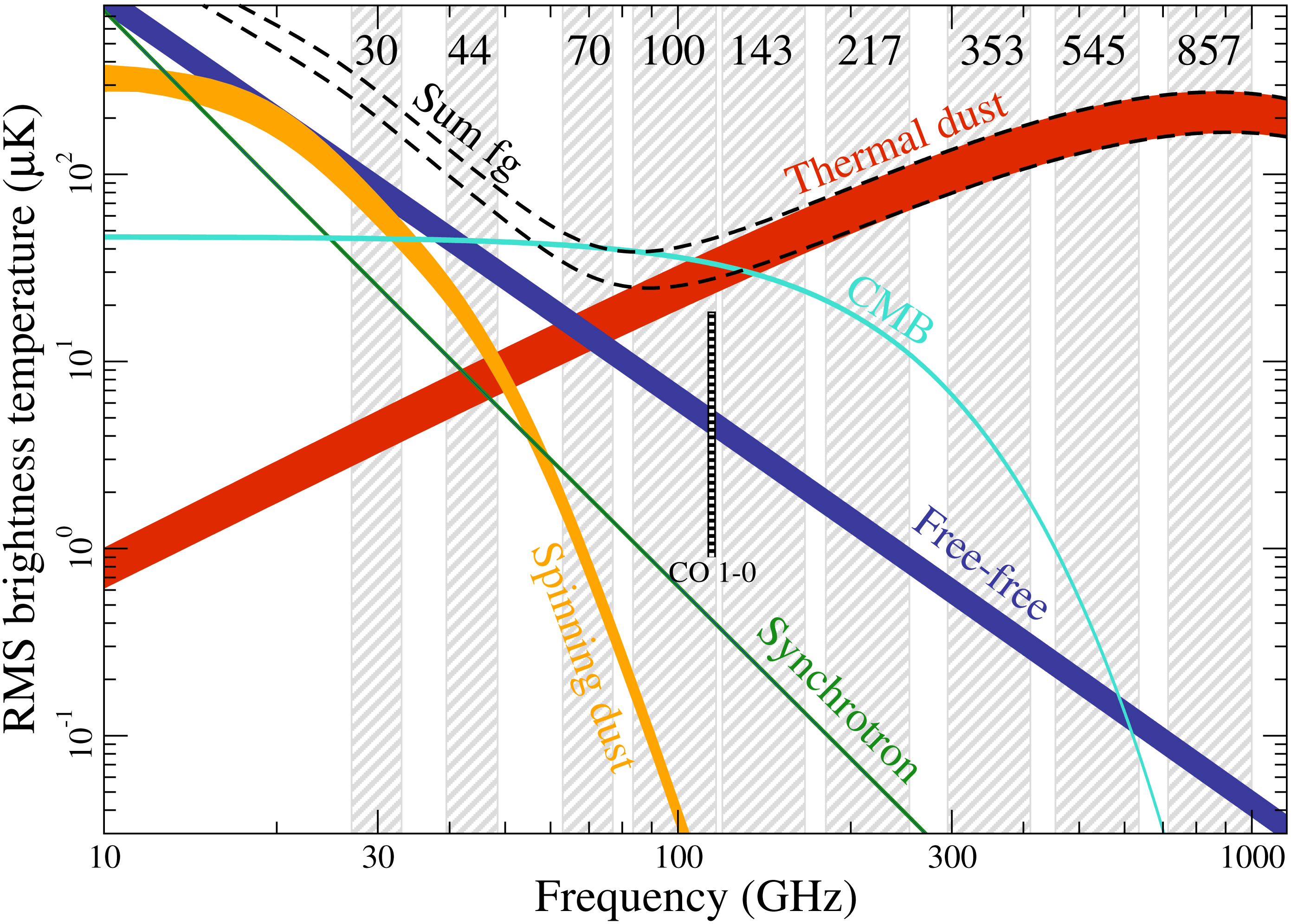}	\includegraphics[width=0.49\textwidth]{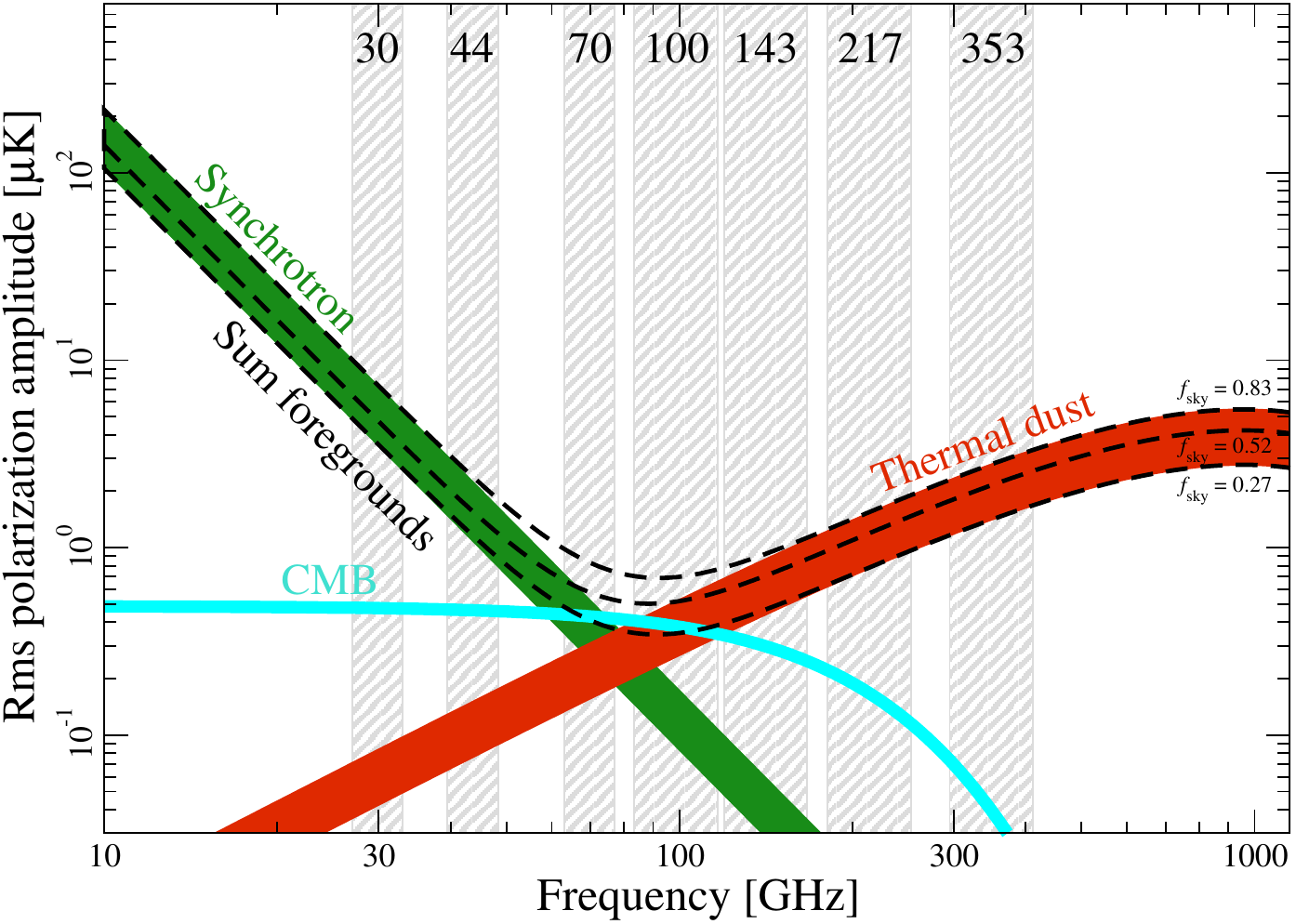}
	}
	\caption{\textbf{Spectral energy distributions of diffuse Galactic foreground emission and the CMB} \textit{Left}: Root mean square (RMS) brightness temperature of the CMB and various foregrounds in microwave and sub-infrared spectrum. (Image from~\citenum{adam2016planck}). \textit{Right}: RMS of polarization intensity. The RMS brightness temperature of the CMB in this case corresponds to the CMB E-mode. (Image from~\citenum{akrami2020planck}). Note that spinning dust may also be weakly polarized. Despite the relative complexity of the temperature foregrounds the CMB signal still dominates near the 100 GHz window, whereas for polarization it never does. }
    \label{fig:foreground_sed}
\end{figure}

The majority of the foregrounds come from our Galaxy, but there is also an important contribution in smaller angular scales coming from extra-galactic point sources. Since the tensor signal is expected on larger angular scales ($\sim1^\circ$), such small-angular-diameter sources are not likely to be relevant to the context of the $B-$mode search. We introduce the relevant foreground components below. The spectral energy distribution of these can be found in figure~\ref{fig:foreground_sed}.

\paragraph{Thermal Dust:} It is due to the heating of dust particles at temperatures ranging from 10--100~K in the interstellar medium. These dust grains are mostly made of graphite, silicates, and polycyclic aromatic hydrocarbons, and their emission can be described by a modified blackbody spectrum~\cite{Ichiki2014}. This emission is polarized, generated by non-spherical grains whose axes align with the magnetic field in the interstellar medium~\cite{Rybicki1979}. According to WMAP 3-year data~\cite{Kogut2007} the fractional polarization is between 1\% in the centre of the Galaxy and 6\% in the Galactic anticenter. At high Galactic latitudes, the polarization is up to 20\%~\cite{PlanckIntXXX2016}.
	
\paragraph{Free-Free:}It arises from the interactions of free electrons in a hydrogen or helium plasma~\cite{Rybicki1979, Draine2010}. At frequencies above a few GHz the brightness temperature spectral index is $\beta=-2.1$~\cite{Dickinson2003}. At higher frequencies, it has been measured to be slightly steeper, $\beta=-2.14$~\cite{PlanckAME2014}. This means it can be dominant near the foreground minimum ($\approx70$ GHz, see left panel of figure~\ref{fig:foreground_sed}). This radiation does not have a net polarization. Whilst emission from a single electron is polarized, the net polarization from an ensemble of electrons with an isotropic velocity distribution is near zero. Therefore free-free is not expected to be a major foreground for the $B$-mode search.


\paragraph{Anomalous Microwave Emission (AME):} It was first identified as an anomalous dust-correlated excess in the microwave range by COBE~\cite{COBE} and by an independent team~\cite{Leitch1997}. The emission mechanism is not yet definitively established~\cite{Dickinson2018}, but as can be seen in figure~\ref{fig:foreground_sed}, it is quite important in intensity in the 8-40~GHz range. It is attributed to electric dipole radiation coming from nanometric grains spinning at several GHz, hence why sometimes is referred to as spinning dust. There is a second possible mechanism that predicts AME should be polarized at the level of a few tens of percent or less depending on the grain chemistry\cite{Dickinson2011}, although polarization measurements over the past two decades have ruled out most such models. Since spinning dust emission is also expected to be weakly polarized ($\lesssim 2-5\,\%$, \cite{Dickinson2011}), a measurement of the AME polarization fraction would provide a decisive test for distinguishing between these mechanisms and determining the origin of the emission. Anomalous emission is omitted from the polarization panel of figure~\ref{fig:foreground_sed} because its polarization fraction remains poorly constrained. Recent measurements report upper limits at the level of $\sim1\%$~\cite{Gonz_lez_Gonz_lez_2025}; this is one argument in favor of an electric dipole origin. 

	
\paragraph{Synchrotron Radiation:}
It is produced by the interaction of relativistic cosmic ray electrons with magnetic fields within the interstellar medium. The primary source of cosmic-ray electrons is supernova remnants~\cite{Rybicki1979}. As electrons spiral around the Galactic magnetic field, they emit radiation. This emission depends on the number and energy spectrum of the cosmic ray electrons and the strength of the magnetic field. The spectrum is approximated by a power-law, with a spectral index in brightness temperature of $\beta\approx-2.7$, exhibiting spatial variations of $\Delta\beta\approx\pm0.2$~\cite{Fuskeland2014}. Synchrotron radiation is intrinsically polarized perpendicular to these field lines. Its spectrum is not yet perfectly understood due to Faraday rotation, which becomes important at frequencies below $\sim$5~GHz~\cite{Lazarian2007}. As discussed in the literature, the limitations and inconsistencies among data sets for synchrotron emission motivate the need for additional surveys, especially in the 10--20~GHz range~\cite{Weiland2022}.


\section{The Canadian Galactic Emission Mapper}
\begin{figure}[b!]
		\centering
		\includegraphics[width=\textwidth]{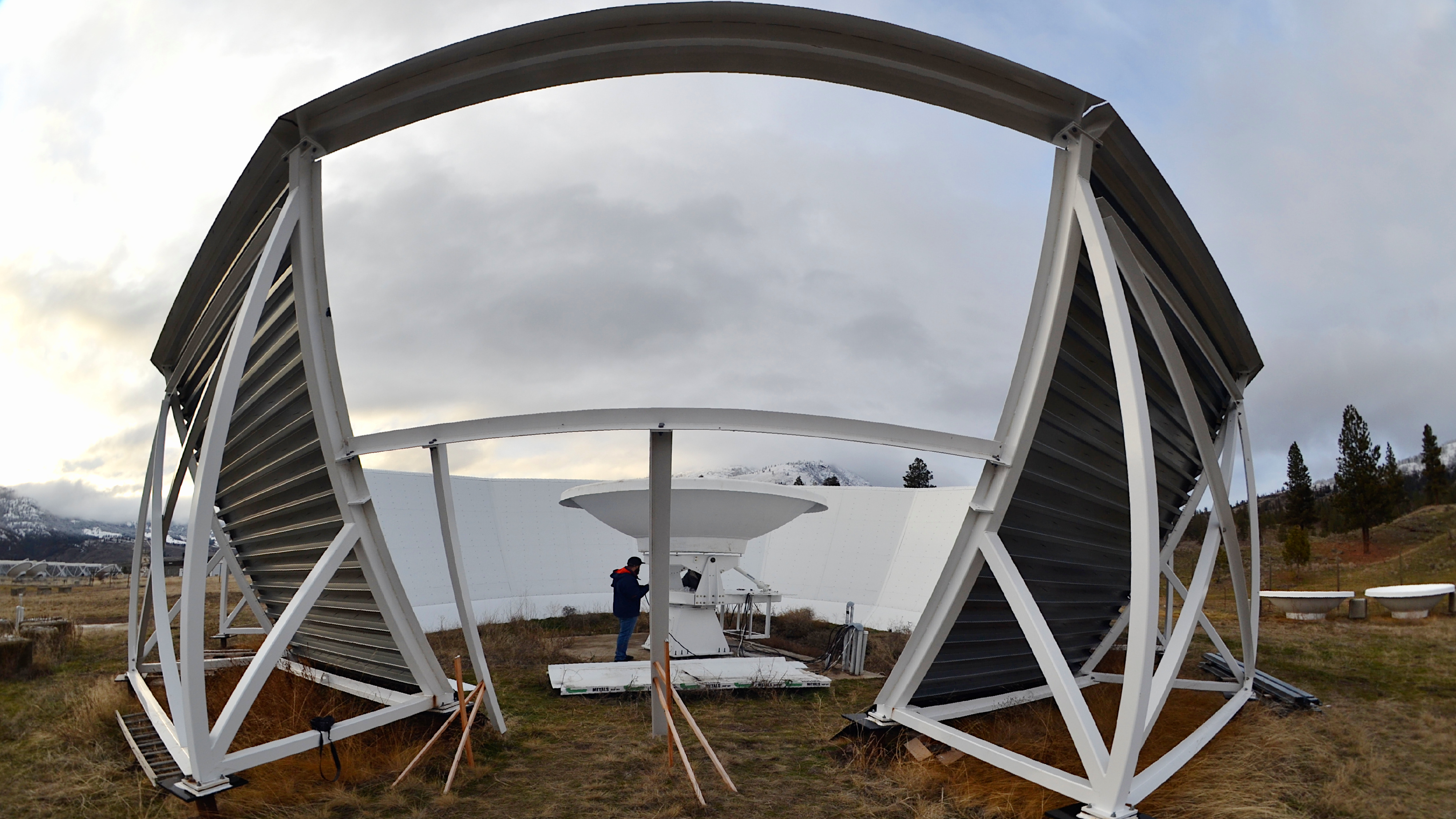}
		\caption{\textbf{CGEM inside its ground screen at DRAO.} The telescope lies inside a ground screen that mitigates signal contamination from the ground and nearby RFI. On the right side of the image, behind the ground screen there are two spare dishes. One of them is intended for a duplicate CGEM proposed to be fielded in Chile.}
		\label{CGEMpic}
\end{figure}

CGEM is a 4-meter single-dish radio telescope located at the Dominion Radio Astrophysical Observatory (DRAO) in Penticton, British Columbia, Canada, see Figure~\ref{CGEMpic}. It has been designed to map polarized Galactic synchrotron emission across the full sky at 8-10 GHz, with the goal of creating high-fidelity low-noise maps of polarized foreground emission for use in $B-$mode searches. Our ultimate goal is to provide full-sky polarization maps in our band. By observing at this frequency range, CGEM measures synchrotron emission with a signal approximately $10^3$ times higher than at CMB observing frequencies ($\sim100$ GHz), while minimizing Faraday rotation effects that contaminate lower-frequency observations~\cite{Ferriere2021}. CGEM will produce high-fidelity templates of polarized synchrotron emission that can then be extrapolated to CMB frequencies, enabling accurate foreground subtraction for the next-generation CMB $B-$mode experiments targeting tensor-to-scalar ratios down to $r \sim 10^{-3}$~\cite{universe8090440}.

\begin{figure}[t!]
	\centering
	\subfigure{
		\includegraphics[width=0.45\textwidth]{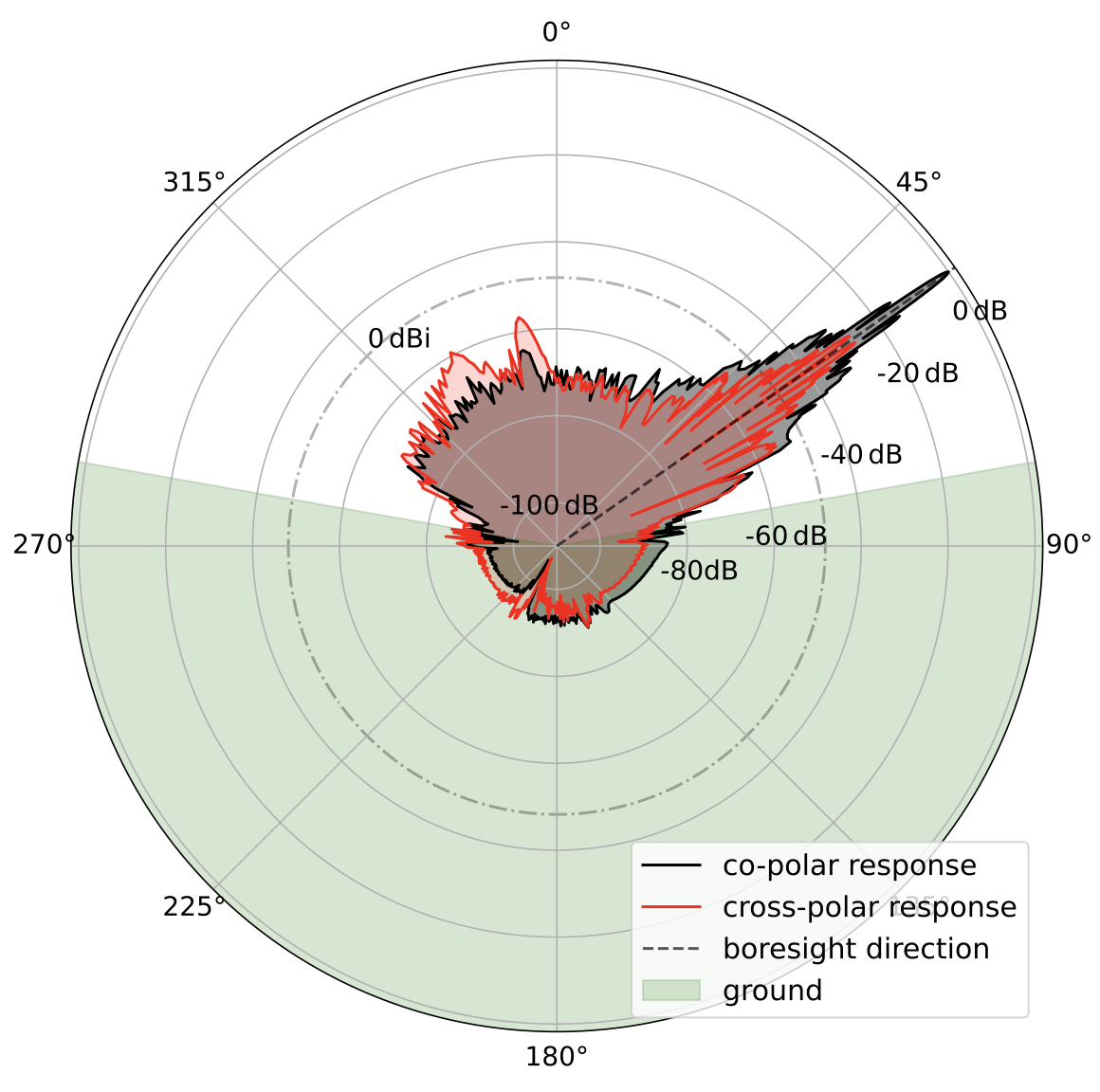}			\includegraphics[width=0.55\textwidth]{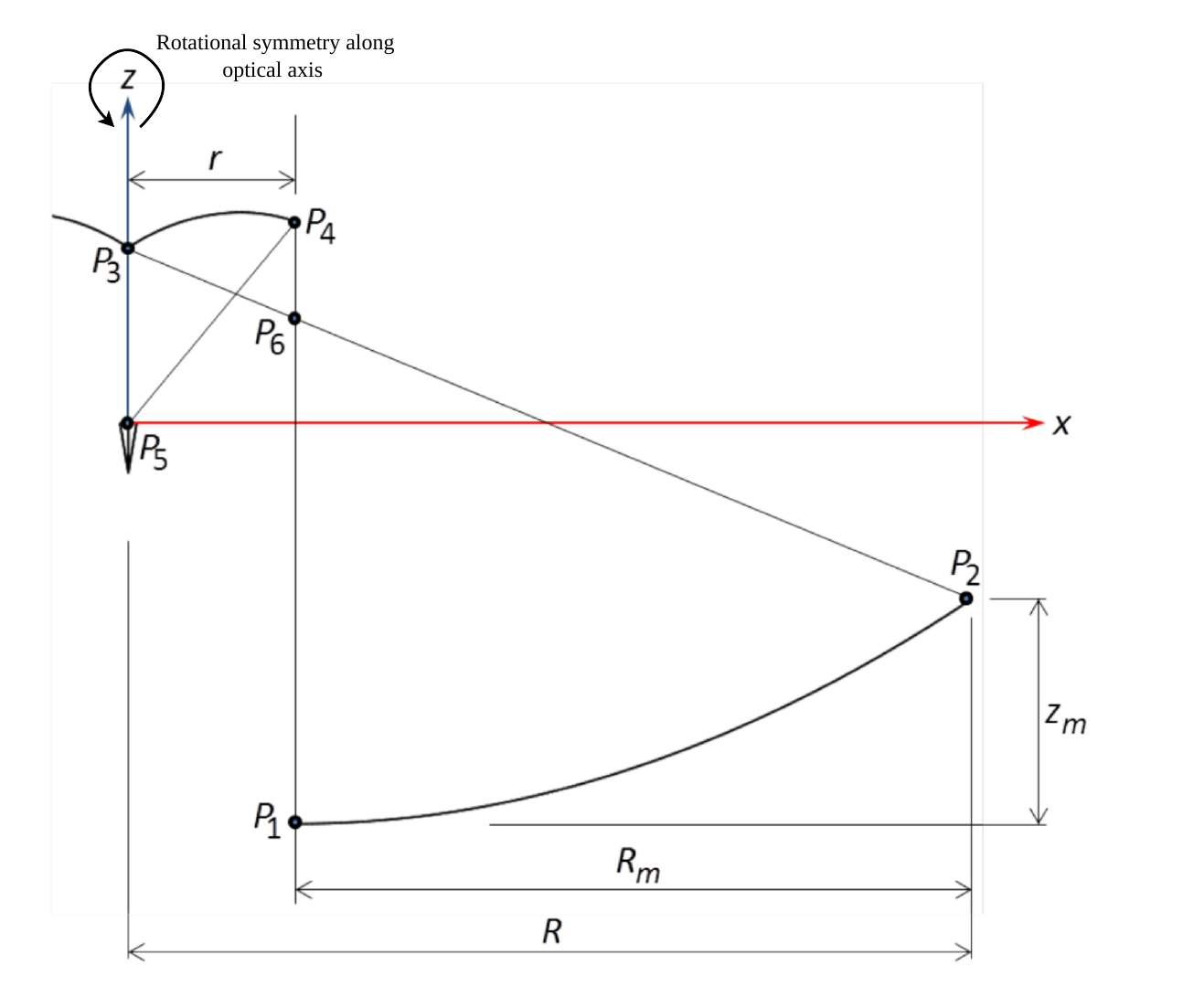}
	}	
	\caption{\textbf{\textit{Left}: Radiation pattern of CGEM primary, secondary (hat), and horn in the presence of ground screen. \textit{Right}: CGEM offset Gregorian optics concept}. $P_1-P_2$ is the CGEM primary which is parabolic section with a focus at $P_6$. $P3-P4$ is the CGEM hat reflector which is part of an ellipse with foci at $P_5$ and $P_6$. Light from parabola (primary) is focused at $P_6$ which is a shared focus between parabola and ellipse. After that the hat feed (ellipse) concentrates the light from locus $P_6$ in $P_5$. A smooth wall horn is placed at $P_5$ to capture the light. Revolving this design around the z-axis creates ring  focus at $P_6$. (Image on the right from~\citenum{TICRA2024}).}
    \label{fig:beam}
\end{figure}

CGEM has been designed with polarization purity in mind. CGEM employs a novel hat-feed \cite{hatfeed_paper}azimuthally-symmetric optical design, see Figure~\ref{fig:beam}.  This eliminates asymmetric feed-leg scattering and minimizes beam-induced polarization systematics. The telescope achieves approximately half-degree angular resolution, sufficient for foreground maps to be used effectively in $B-$mode detection at angular scales relevant for cosmology ($\ell \lesssim 100$, $\theta \gtrsim 1.8^\circ$). The receiver has a spectral resolution of 1 MHz, which allows us to surgically excise radio frequency interference (RFI) and to generate modest resolution subband maps with our own data. CGEM captures circularly polarized light, that is subsequently separated into two orthogonal linear polarizations, allowing us to measure Stokes Q and U, which characterize the direction and magnitude of linear polarization across the sky.

\begin{figure}[b!]
		\centering
		\includegraphics[width=0.7\textwidth]{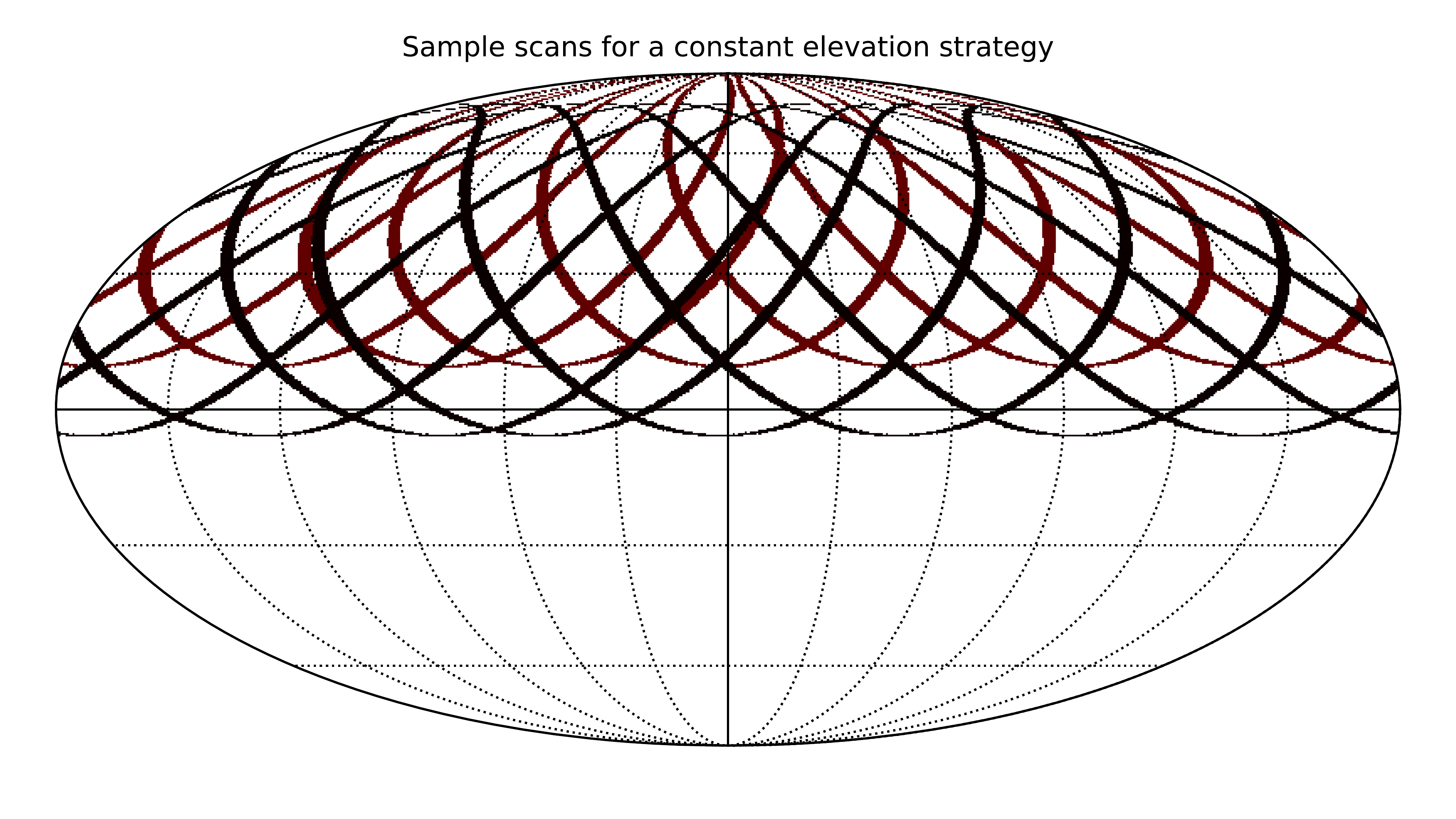}
		\caption{\textbf{Simulated scans in celestial coordinates}. Each of these bands shows fifteen minutes of scanning every three hours as the sky drifts through the field of view. Note how in the 40$^\circ$ case (red) CGEM crosses the celestial north pole once per scan. In the 55$^\circ$ scan (black) CGEM crosses the equator, achieving a better overlap with CMB experiments observing in the southern hemisphere.}
		\label{SampleScanStrategy}
	\end{figure}
The goal of achieving polarization purity has been kept in mind in each decision we took during the design of the instrument, while ensuring we meet the sensitivity necessary to be useful for the next generation of CMB instruments~\cite{Fuskeland2023, CMBS42022}. The design choices motivated by this goal are:
	\begin{itemize}
		\item \textbf{Polarization purity:} CGEM employs an azimuthally-symmetric hat-feed optical design, which is an offset Gregorian system rotated about the optical axis, see Figure~\ref{fig:beam}. The design of the optics was optimized to minimize polarization artifacts in our maps following the novel approach from~\citenum{MacEachern2023}.
		\item \textbf{Detecting polarization in the circular basis:} Correlating in the circular basis produces $Q$
		and $U$ directly from the cross-correlation of two channels, which is relatively insensitive to gain fluctuations in either channel individually.
		\item \textbf{Angular resolution:} B-mode signals are expected on angular scales large enough that a resolution of $\sim0.5^\circ$ is sufficient.
		\item \textbf{Scan strategy:}To mitigate $1/f$ noise from the atmosphere, we scan at constant elevation in azimuth~\cite{Delabrouille1998}, see figure~\ref{SampleScanStrategy}.
	\end{itemize}
In its final configuration, CGEM will have a cryogenically cooled feed horn, followed by an orthomode transducer (OMT), which separates right-hand and left-hand circular polarizations into two channels. After this, the two polarized signals go through cryogenically cooled low-noise amplifiers (LNAs). Following this, the signal goes through the analog chain, a double channel single-sideband superheterodyne receiver in which the signals get mixed down from 8-10 GHz to 0-2 GHz, low-pass filtered and amplified, before being digitized by readout electronics, which sample the signal at 4 GS/s. The signal is subsequently processed in the frequency domain in the FPGA fabric. 

    \begin{figure}[b!]
		\centering
		\includegraphics[width=1\textwidth]{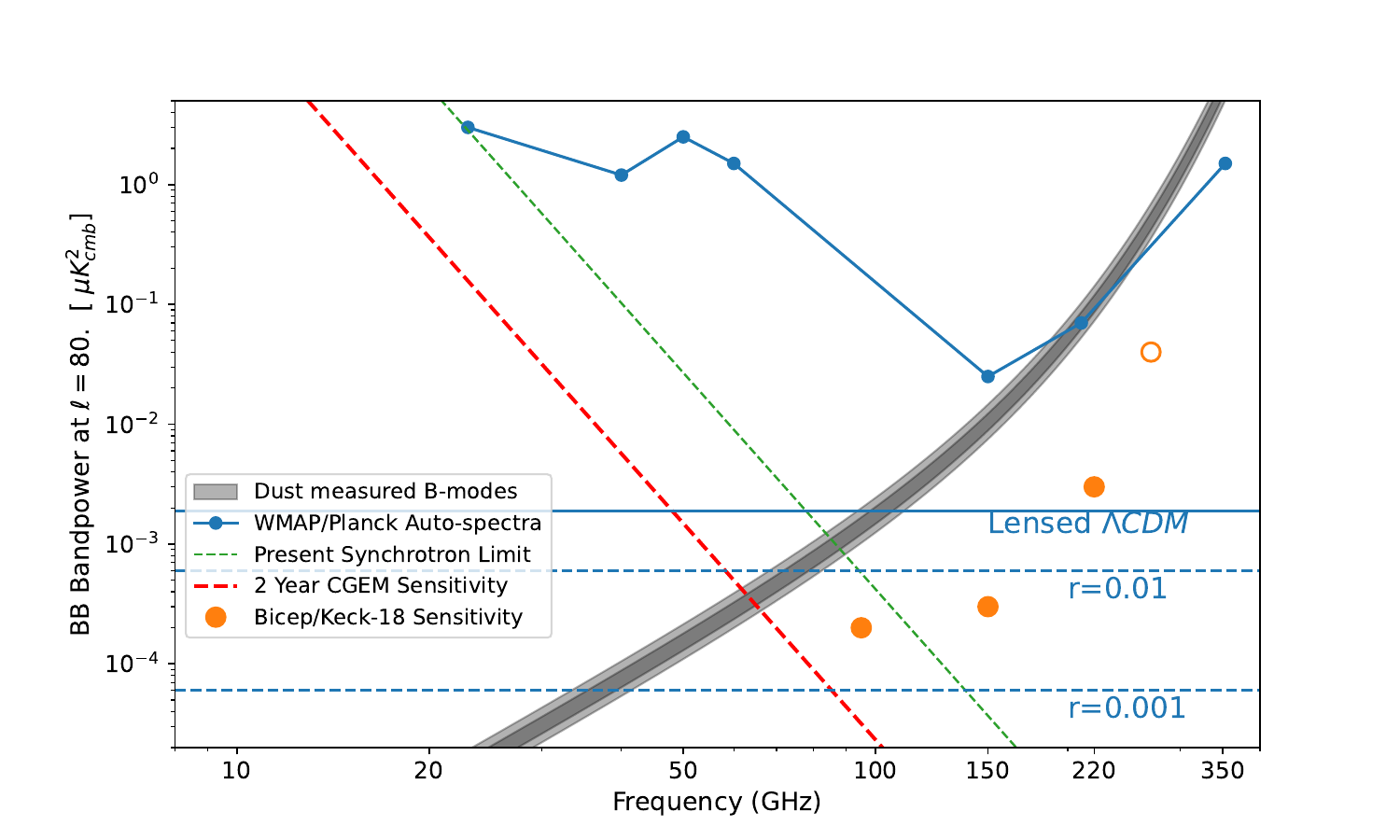}
		\caption{\textbf{Expectation values and noise uncertainties of the $\ell\sim80$ BB band-power in the BICEP/Keck field.} The solid and dashed black lines show the expected signal power of lensed-$\Lambda$CDM and $r_{0.05}$ = 0.01. Since CMB units are used, the levels corresponding to these are flat with frequency. The blue bands show the 1 and 2$\sigma$ ranges of dust, and the red shaded region shows the 95\% upper limit on synchrotron in the baseline analysis including the uncertainties in the amplitude and frequency spectral index parameters (Async, 23, $\beta_s$ and Ad, 353, $\beta_d$). The BICEP/Keck auto-spectrum noise uncertainties are shown as large blue circles, and the noise uncertainties of the used WMAP/Planck single-frequency spectra evaluated in the BICEP/Keck field are shown in black. The blue crosses show the noise uncertainty of selected crossspectra, and are plotted at horizontal positions such that they can be compared vertically with the dust and sync curves. The CGEM 2 year sensitivity curve is derived from equation~\ref{eq:1.38}. Image after~\citenum{Ade_2021}.}
		\label{fig:BBpower}
	\end{figure}

To be useful in the search for $B$-modes, CGEM's sensitivity should be sufficient that the limiting factor in our maps is the complexity of the foreground emission, not instrumental noise. At present, and for the next few years, the BICEP experiment will have the best sensitivity on the sky, with a forecast linear polarization sensitivity of $\sim$2~$\mu$K-arcmin~\cite{BICEPArray2018}, see Figure~\ref{fig:BBpower}, which is equivalent to an RMS sensitivity per square degree of $\sigma_{95\,\mathrm{GHz}} = 16\,\mathrm{nK_{CMB}\,deg^{-2}}$. For CGEM to obtain an equivalent sensitivity at 95\,GHz, we must have a 9\,GHz sensitivity of
\begin{align}
  \sigma_{9\,\mathrm{GHz}}
    &= \left(\frac{\Delta T_B}{\Delta T_{\mathrm{CMB}}}\right)_{95\,\mathrm{GHz}}
       \left(\frac{9\,\mathrm{GHz}}{95\,\mathrm{GHz}}\right)^{-2.7}
       \sigma_{95\,\mathrm{GHz}} \notag \\
    &= \left(0.796\,\frac{\mathrm{nK}}{\mathrm{nK_{CMB}}}\right)
       \times 579.961
       \times 16\,\mathrm{nK_{CMB}\,deg^{-2}} \notag \\
    &= 7.388\,\mu\mathrm{K\,deg^{-2}}, \label{eq:1.38}
\end{align}
where we have used the unpolarized synchrotron scaling relation to derive a conservative minimum sensitivity. From this consideration we could use the radiometer equation\cite{10.1063/1.1770483},
\begin{equation}
  \sigma = \frac{T_{\mathrm{sys}}}{\sqrt{B\tau}}, \label{eq:1.39}
\end{equation}
to determine the necessary integration time, where $\sigma$ is the rms sensitivity after an integration time of $\tau$, over a
frequency bandwidth $B$, using an instrument with a system temperature of $T_{\mathrm{sys}}$. Plugging in $\sigma_{9\,\mathrm{GHz}}$
and the parameters for CGEM ($B = 2\,\mathrm{GHz}$, $T_{\mathrm{sys}} = 10\,\mathrm{K}$)
and solving for $\tau$, we find $\tau \simeq 916\,\mathrm{s\,deg^{-2}}$. With
$20{,}600\,\mathrm{deg^2}$ in the Northern sky, this means \textsc{cgem} will have to observe
for ${\sim}218$ days to achieve the required sensitivity. However, the Sun's bright emission
corrupts daytime polarization measurements and so only nighttime observations can be used for
CMB science. This means CGEM will need ${\sim}436$ days worth of nighttime data to achieve its required sensitivity. 

Other experiments such as QUIJOTE~\cite{RubinoMartin2023}, S-PASS~\cite{Carretti2019} or C-BASS~\cite{King2010} have a similar goal: the study of polarized Galactic foregrounds. CGEM is an addition to these surveys, with some characteristics that make it unique: it is a purpose-built polarimeter, which enables surgically extracting RFI from satellites, among other sources, and should not be affected significantly by Faraday rotation. Once commissioning of the cryogenic system is complete, we propose to build a copy of CGEM in the Southern Hemisphere, likely located near Concepción, Chile. Our ultimate goal is to provide a full-sky polarization map in the 8-10 GHz band. 

\section{Orthomode Transducer}
\begin{figure}[t!]
	\centering
	\includegraphics[width=0.5\textwidth]{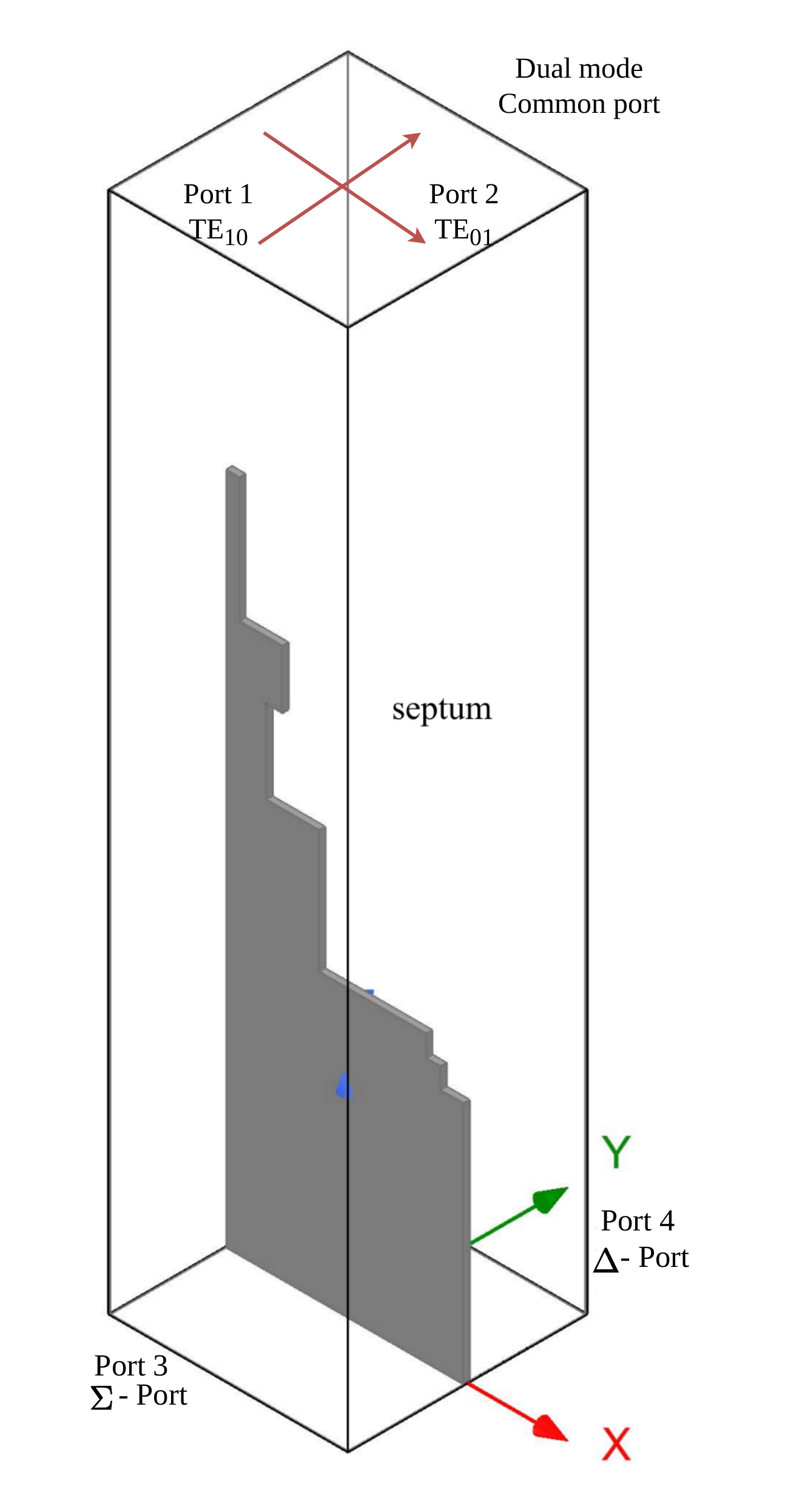}
	\caption{\textbf{CGEM's septum polarizer and its port definitions.} Note that the X-axis is along the direction of septum, the Y-axis is orthogonal to the septum, and Z-axis is the optic axis. The septum polarizer is located at X-Z plane. The CGEM septum polarizer has six sections. Each section can be viewed as a ridge waveguide, in which the length and height is chosen to equally divide the guided modes $E_x$ and $E_y$, introduced into the common-port and convert to left and right polarized signals $\frac{E_x \pm iE_y}{\sqrt{2}}$in the output ports. 
    }
    \label{septum_ports}
\end{figure}

In the CGEM optical design every element up to the polarizer (primary mirror, hat reflector, horn, and circular waveguide) has continuous rotational symmetry. The septum polarizer is the first element that breaks the symmetry and discriminates between the two incoming circular polarizations. A septum polarizer is a type of orthomode transducer (OMT). CGEM's septum polarizer has four ports which are shown in Figure \ref{septum_ports}. There are two modes at the common port which are distinguishable by their linear polarization states $TE_{10}$ and $TE_{01}$ and one in each single mode rectangular waveguide port. There is a one to one relationship between modes and ports. The device is symmetric about the plane of the septum polarizer. The septum polarizer is comprised of six sections, in which each of them can be viewed as a unique delay section. The delay sections of the septum polarizer are designed such that the incident radiation at the common port (i.e., ports 1 and 2 as depicted in Figure~\ref{septum_ports}) is separated by circularly polarization state at the output ports (i.e., $\Sigma$ and $\Delta$ ports 3 and 4). 

Each section of the septum polarizer can be looked at as a single ridge waveguide which supports an electric (E-)field along and orthogonal to the ridge. The cutoff frequency of the E-field along the ridge is much lower than a waveguide with the same cross section. This mode also has a lower characteristic impedance, which leads to the alignment of the field with this specific mode \cite{balanis2012advanced}. 

\begin{figure}[t!]
	\centering
    \includegraphics[width=1\textwidth]{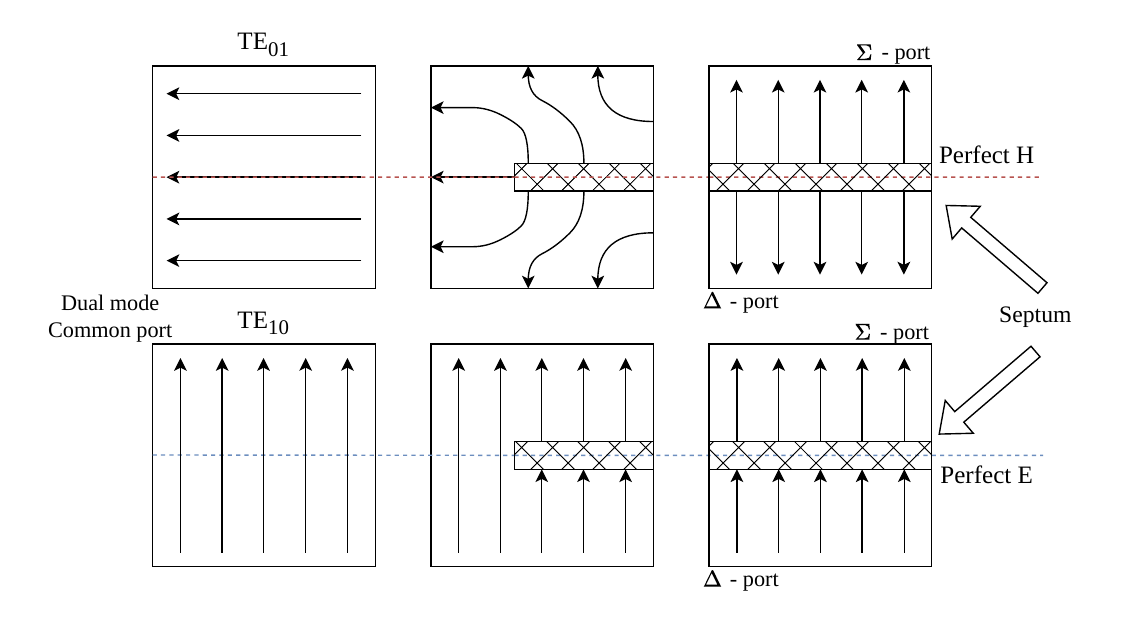}
    \caption{\textbf{Effect of the septum on degenerate modes within the square waveguide.} Three cross sections are shown, at the input, midway and at the output of the polarizer. The TE$_{10}$ mode is orthogonal to the septum and goes unchanged as it travels through various sections of septum polarizer. In this case the septum polarizer acts as a perfect E-plane. TE$_{01}$ is parallel to septum and rotates by 90 degrees as it passes through the septum. Due to symmetry each side of the septum rotates the field in opposite directions. In this case the septum acts as a perfect H-plane. The field amplitude of each mode is equally divided and phased along the septum to produce left and right circular polarization at 2:1 rectangular output waveguide ports. Notice from the patterns described above, if the input singal is circularly polarized, the signal will end up at either $\Sigma$ or $\Delta$  ports as a linear polarization.}
    \label{fig_OMT_modes}
\end{figure}

As shown in Figure~\ref{fig_OMT_modes}, in the single ridge waveguide the following relationships exist between modes parallel to the ridge and orthogonal to the ridge: $\beta_{g \;\parallel} > $  $\beta_{g\; \perp}$, $Z_{0 \;\parallel} < $  $Z_{0 \;\perp}$ and $\lambda_{g\;\parallel} < $  $\lambda_{g\;\perp}$. This means that for a given physical length, the electrical length of the mode parallel to the septum is longer than the orthogonal mode. The differences in length over the six steps are optimized such that the degenerate modes with a 90$^\circ$ phase shift sent at the common port arrive at port 3 or 4 (depending on polarization) in phase.
A polarizer can be benchmarked by: 
\begin{itemize}
	\item Its quality of impedance match at all electrical ports;
	\item The cross polarization discrimination between two polarized input signals; and 
	\item The isolation between ports at common port and the other two ports (1 $\leftrightarrow$ 2, 3 $\leftrightarrow$ 4). 
\end{itemize}

At any given discontinuity between two sections of the waveguides, mode matching techniques need to be used to determine the coupling of the incident mode to higher modes. Using these methods becomes extremely important if the geometry allows a higher order mode to propagate within the operating frequency band. Additionally, if a discontinuity occurs at the boundary of two waveguides in which one or both guides support the degenerate mode within the operating frequency, mode matching can be a powerful tool to determine and refine the coupling between degenerate modes \cite{wghandbook}.

\begin{figure}[t!]
	\centering
	\includegraphics[width=0.85\textwidth]{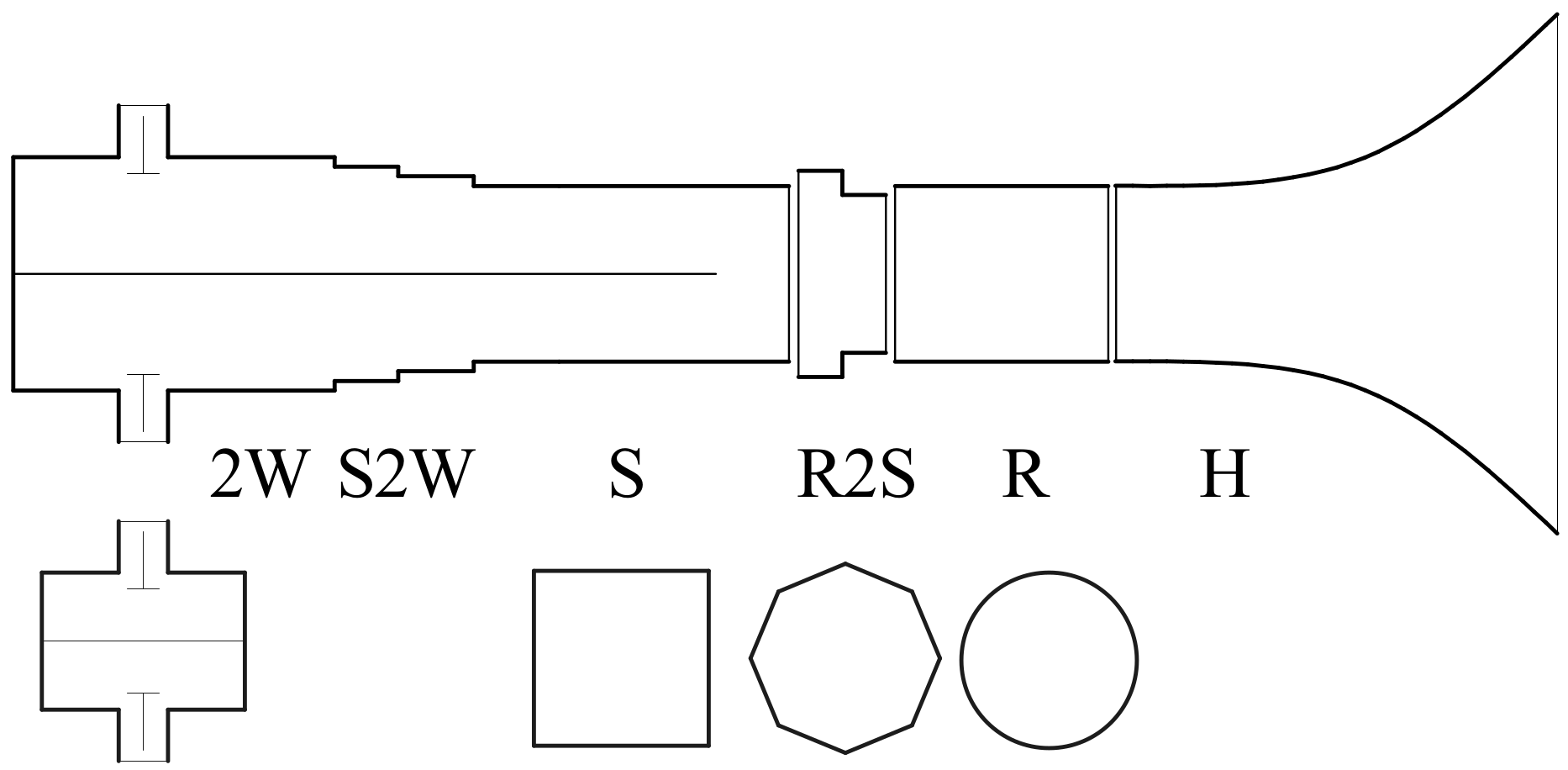}
	\caption{\textbf{Sketch of the optic tubes cross section.} The received light arrives at the circular symmetric horn (H). A custom circular waveguide (R) guides the incoming waves towards the round to square transition (R2S). The R2S is comprised of two concentric octagons. The R2S acts as quarter wave transformer. The square waveguide (S) houses the septum polarizer which separates the circularly polarized light into two separate guides. The separated polarized light at the end of the septum polarizer are housed at the custom rectangular waveguide and goes through a three step inhomogeneous transition from the custom rectangular guide to WR90 waveguide (S2W). The separated polarized light in two WR90 waveguide housing (2W) are terminated with a K connector transformer which guides the wave toward the radiometer.    
    } \label{CGEM_optic_tube}
\end{figure}

The septum polarizer is housed in the optic tube, see Figure \ref{CGEM_optic_tube}. Within the optic tube mode launchers on either end of the septum polarizer are housed in WR90 waveguides. They launch a linear polarization wave inside the waveguide. Note that as the two mode launchers are located mirror to the separator, their radiated electric fields have a 180-degree phase shift with respect to each other. Once the wave has been launched in the WR90 waveguide, an inhomogeneous three-step transition changes the cross section of WR90 waveguide to another custom rectangular waveguide. This intermediate rectangular waveguide dimension is chosen to have the same cutoff frequency as the optics. The cutoff frequency of 7.7 GHz is associated with the square side length. CGEM's cutoff frequency is chosen such that it only allows for the first mode of operation within circular/square waveguide and polarizer given the operating bandwidth of 8--10 GHz. The impedance of the waveguide for a given mode is a function of frequency and the cross section geometry. In contrast, the characteristic impedance presented by the mode launcher's coaxial lines are fixed at 50 $\Omega$. The stepped transition works as an impedance transformer in between the mode launcher and polarizer (matching circuit) with maximum power throughput.

 Up to this point the two channels are completely isolated in the housing by the separator and only TE$_{10}$ modes will propagate. Once TE$_{10}$ mode has passed the separator, the waveguide polarizer septum rotates the TE$_{10}$ in each channel. This is such that after passing through the polarizer, half of the TE$_{10}$ power is transformed into TE$_{01}$ with a 90-degree phase shift (so that it becomes parallel to septum). It is important to note that the initial 180-degree phase shift between the electric fields at two different WR90 channels are orthogonal to the polarizer. The polarizer rotates half the power in each channel from orthogonal to parallel with respect to the septum (toward one direction). The rotation of fields on each side are opposite of each other which results in creating right-hand (RHCP) and left-hand (LHCP) circular polarization.

  \begin{figure}[b!]
	\centering
	\includegraphics[width=0.74\textwidth]{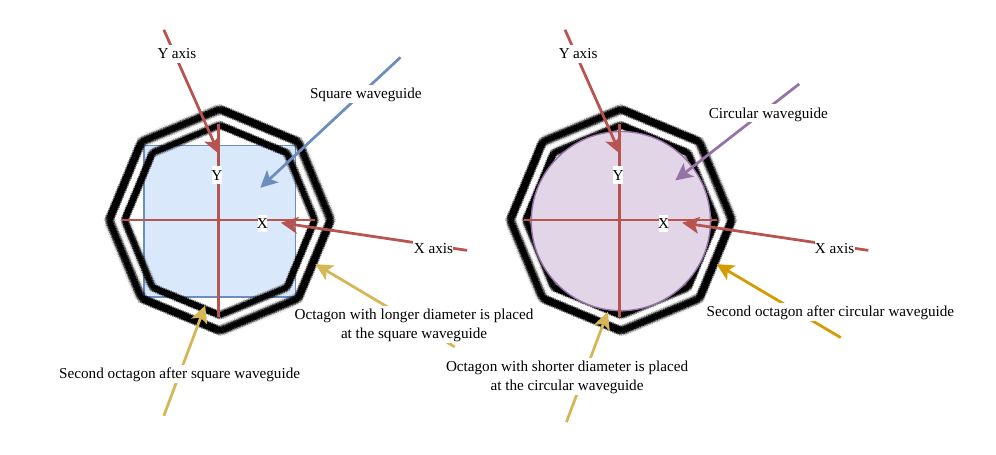}
	\caption[optic tube parts]{\textbf{The two octagons serve as a transition between the square to circular waveguides sections.} The polarizer sits in the Y-axis further down in the square waveguide. Moving along optic axis toward sky from square waveguide, the octagon with longer diameter is placed after the square waveguide. The longer diameter octagon followed by the shorter diameter octagon and ends with the circular waveguide. The two octagons act as a stepped quarter wavelength circular to square waveguide transition. 
    }\label{circular to square transition}
\end{figure}

At this point the circularly polarized signals are propagating in the square waveguide. A pair of back to back, concentric and aligned octagons are used to connect the square waveguide to the circular waveguide. This is referred to as the circular to square transition parts. Note that these two octagons are inhomogeneous\cite{matthaei1980emt} and have a different cut off frequency compare to the square and circular waveguides. The phase delay of the octagon pair acts as quarter wave transformer for mode and impedance matching between square and circular waveguide. The square and circular waveguide are designed to have the same cutoff frequency of 7.7 GHz. The octagons are placed in the square waveguide such that the two orthogonal long diagonals of the octagon (which are twice the length of the circumradius) align themselves to the two orthogonal degenerate linear polarization axes.

The two orthognal RHCP and LHCP polarized signals in the circular waveguide are launched to space using a smooth wall horn. The smooth wall horn follows a design that matches the frequency dependent impedance of the optics tube to the wave impedance of free-space, which is $\sim$377 $\Omega$ per square. This is described by looking at the polarizer as a transmitter of RHCP or LHCP signals. Using the reciprocity theorem one can see that the polarizer separates RHCP and LHCP to linear polarizations.

\section{Coherent radiometer}

\begin{figure}[t!]
	\centering
	\includegraphics[width=1\textwidth]{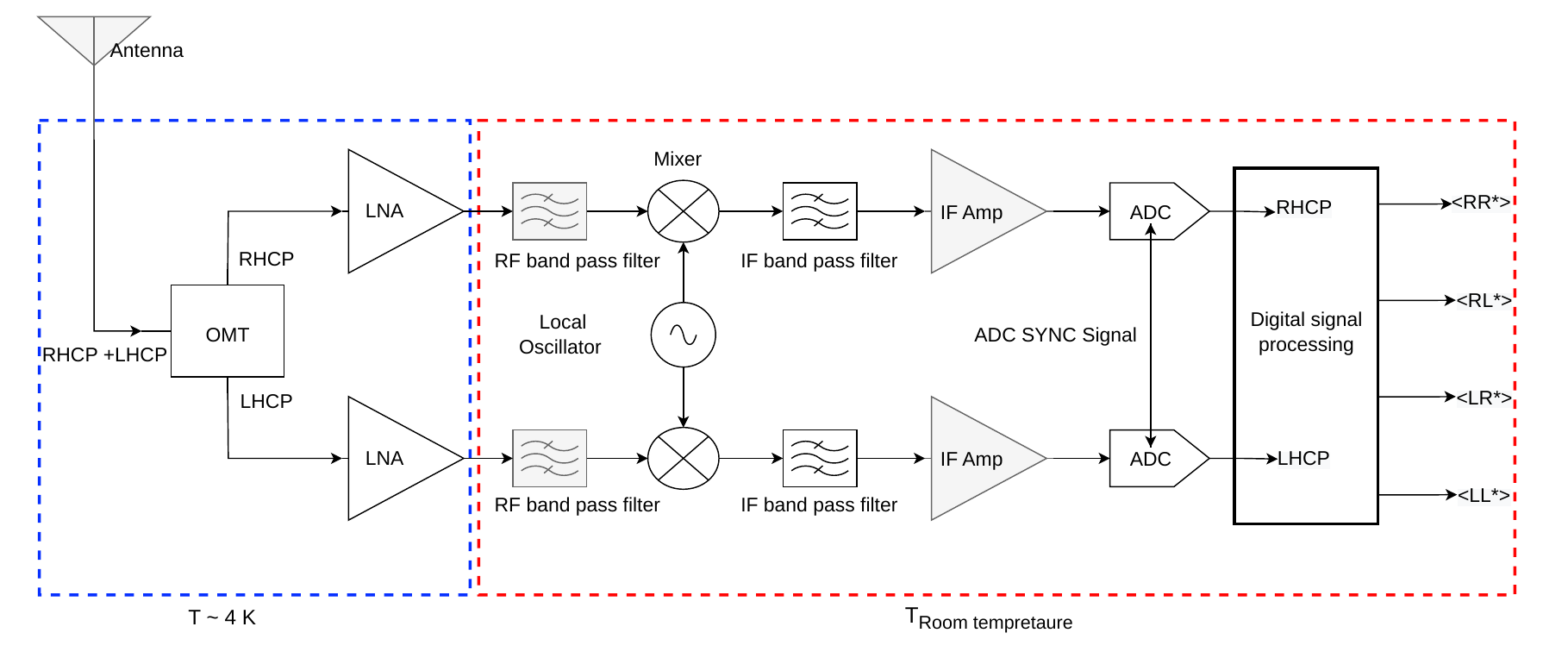}
	\caption{\textbf{CGEM receiver block diagram}. The front-end componenets are cooled to 4 K. The warm electronics down-convert the signals coherently from 8-10 GHz to 0-2 GHz. The filters ahead of the mixer block unwanted signals while the filter following the mixers prevent aliasing in the digitization stage. Note that The E fields at OMT outputs propagate through coaxial waveguides that operates in TEM mode. There is a one to one relationship between E-field and the voltage in this mode. Therefore the electric fields at the output ports of the OMT can be looked at as a voltages.  }\label{CGEM_analog_chain}
\end{figure}

The RHCP and LHCP electric fields are separated into the two OMT output ports. The OMT outputs are connected to a double channel superheterodyne receiver as shown in Figure~\ref{CGEM_analog_chain}. To reduce the noise temperature of the LNAs, they are cooled down to 4 K. At the output of the LNAs the signal to noise ratio is greater than one, therefore the rest of the system operates at room temperature. At the output of the LNAs, a selective band pass filter rejects the signals oustide the 8--10 GHz band. At this point a mixer with an injected 8 GHz tone from a local oscillator down-converts the signal to the 0-2 GHz baseband. Finally, intermediate-frequency (IF) amplifiers further amplify the signal and send it to two synchronized analog to digital converters. The digital system computes the auto- and cross-correlation products. The LNAs are chosen to be broadband so that their noise bias does not affect cross-correlation products. Since CGEM is correlating in the circular basis, the correlation products might be expressed in terms of
\begin{align}
	&\langle RR^* (t) \rangle \; = \; T(t) + V(t) + \text{Cross polar terms + LNAs noise bias}, \\ 	&\langle LL^* (t)\rangle \; = \; T(t) - V(t) + \text{Cross polar terms + LNAs noise bias},
	\\ 	&\langle RL^* (t)\rangle \; = \; -Q(t) + iU(t) + \text{Cross polar terms},
	\\&\langle	LR^* (t)\rangle \; = \; -Q(t) -	 iU(t) + \text{Cross polar terms}.	
\end{align}
In the left hand side of the above equations, $RR^{*}$, $LL^{*}$, $RL^{*}$, and $LR^{*}$ are respectively the auto correlations of right and left circularly polarized signals followed by the cross correlation products of the polarized signals. A block diagram of the CGEM signal chain is shown in Figure~\ref{CGEM_analog_chain}. CGEM has double balanced passive mixers which down-convert the upper side band and lower side band of the RF frequency equally. The presence of band pass filters are necessary to reduce the leakage from the RF lower side band at 6-8 GHz to down-converted IF signals. The low-pass filter on the IF side keeps all the signals within the first Nyquist zone of the analog-to-digital converters (ADCs) and suppresses aliasing. As CGEM's science goal is to measure the linear polarization of galactic foregrounds, tracking the phase across each radiometer channel is an extremely important calibration task. The radiometer parts were chosen such that:
\begin{itemize}
    \item The matching network between components is maximized;
    \item The noise at each stage is small compared to the signal from previous stages (except at the LNAs);
    \item The overall gain of the instrument provides enough power to overcome the inherent noise of the ADCs. This gain is low enough that RFI does not frequently saturate the system;
    \item All the active components operate linearly, orders of magnitude below their saturation point;
    \item The radiometer has component symmetry between each channel; and
    \item The analog chain is temperature controlled for better gain and phase stability.
\end{itemize}

\section{digital spectrometer}\label{sec:readout}
\begin{figure}[b!]
	\centering
	\includegraphics[width=1\textwidth]{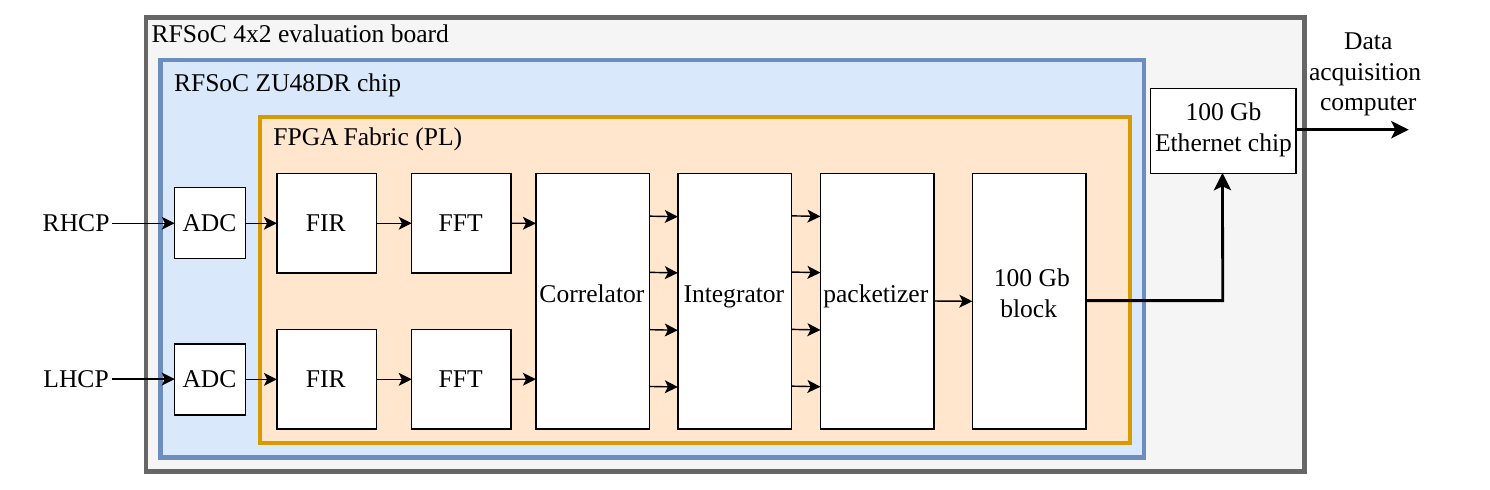}
	\caption{\textbf{Block diagram of the real-time digital spectrometer implemented in the FPGA fabric.} CGEM incorporates RFSoC ZU48DR chip for its real-time processing pipeline. This chip is commercially packaged and made available through RFSoC4x2 evaluation board. The FIR, FFT and 100 Gb blocks are implemented by CASPER\cite{CASPER}.}\label{Digital backend}
\end{figure}

This section describes the digital spectrometer for CGEM. The two RHCP and LHCP signals are terminated into two high-speed RF Analog-to-digital Converters (ADCs) and critically sampled at 4096 MS/s. The ADC sampling rate allows for 2048 MHz analog bandwidth within the first Nyquist zone. Once digitized, the data goes through a signal processing pipeline, shown in Figure \ref{Digital backend}, and then streamed through a 100 GbE link to a data-acquisition PC. The digital backend uses an RFSoC 4x2 Evaluation platform that offers four 14-bit RF ADCs, Field-Programmable Gate Array (FPGA) fabric, and a 100GbE interface. The signal processing pipeline is implemented in the programmable fabric and consists of a finite-impulse-response (FIR) filter, frequency-domain conversion, correlation computation, integrate for a user-specified periods of time and then packetizing the data to be sent over a 100GbE link to the data-acquisition computer. 

Every 4096 samples (equivalent to 1 $\mu\text{sec}$) of data goes through an FIR filter. The FIR is a weighted moving average filter which has the purpose of de-correlating the frequency bins. The time-filtered voltages are transformed to frequency domain in the fast Fourier transform (FFT) block with bin resolution of 1 MHz. As real FFT algorithms are incorporated, all the frequency data is symmetric around DC point. Therefore, to reduce the output data rate all the frequency bins with negative frequency are dropped at this stage. In the next stage, for every frequency bin, the correlator engine calculates the auto- and cross- correlation products separately. The output of correlator are the auto-correlation and cross-correlation power spectra. This output power corresponds to white noise in the frequency domain so it can be averaged out for periods of time for individual frequency bins. In order to reduce the data rate and noise level, multiple spectra are co-added in the integrator. The integrator block is parameterized and can be programmed to add multiple spectra anywhere in the range of $10 - 10^7$ spectra. 

Finally, the packetizer block, serializes the coadded data, synchronizes it with the system clock, and sends the data to a data acquisition computer with streaming user datagram protocol (UDP) algorithm via fiber optic cables. The packetizer block performs upto 50 Gbit/s. The packetizer block is also equipped with a trigger function which simultaneously captures a 32 $\mu \text{sec}$ burst of latest ADC data, assemble them into TCP packets and send them to a 1GbE housekeeping interface. This is a powerful tool in terms of observability into the digital signal processing pipeline and troubleshooting the instrument. The correlator data is repackaged at the data acquisition computer and sent out to a High-performance computing cluster for scientific analysis.

\section{First results}
CGEM's commissioning started in February 2025. Since then, we have collected around 4,300 hours of data with a warm radiometer, operating continuously since December 2025. In section~\ref{sec:datared} we go over the data reduction pipeline we have built to produce the first Stokes I maps from CGEM. We describe our map-making pipeline and show these first maps in section~\ref{sec:maps}.

\subsection{Data reduction \& calibration}\label{sec:datared}
As described in section~\ref{sec:readout} CGEM samples the sky at 1 MHz resolution at the cadence of 200 Hz (5 ms). The nominal scan speed of CGEM is 2 rpm, which corresponds to a beam smearing of less than 0.5\% of our beam width ($\sim0.5^\circ$). CGEM produces $0.5$ TB of data daily. This data needs to be processed and reduced in order to enter our science pipeline. The first step is to produce a kurtosis-based RFI mask \cite{nita2007radio,taylor2019spectral} based on auto correlation data, which is subsequently applied to our four correlation products. This filter exploits the fact that thermal noise obeys Gaussian statistics, while RFI typically produces non-Gaussian signals with excess kurtosis. The fraction of excised data is typically below $\sim0.5\%$, demonstrating how useful our FPGA engine is at surgically removing RFI.

After RFI flagging, we convert the data to brightness temperature units by multiplying by a calibration factor based on astrophysical sources. Our current approach is to map source transits while astronomical sources gradually rise or set across the path of the CGEM scan strategy. By extracting calibrator sources from these transit event maps through aperture photometry, Gaussian fitting or point-spread function (PSF) fitting and comparing them to the source flux models provided by~\citenum{Weiland_2011}, we are able to obtain and validate estimates of the channel gains over time. We are currently working on expanding this framework and producing a time-dependent gain model to correct for thermal drifts of the amplifiers, but for the meantime assume a flat gain profile with time.

After converting the data to brightness temperature units we produce eight band averages of 250 MHz each. Once we have these products, we manually flag and remove data affected by bad weather or contaminated by lunar or solar emission. Finally, instrumental drifts are removed by a high-pass filter. 
\subsection{First maps}\label{sec:maps}
We follow a maximum-likelihood approach to the map-making problem~\cite{Tegmark1997}. We model the time-ordered data (TOD) at sample $i$ as
\begin{equation}
	d_i = \sum_{p=1}^{N_{\mathrm pix}} A_{ip} \, m_p + \sum_{k=1}^{N_{\mathrm az}} G_{ik} \, g_k + n_i,
	\label{eq:data_model}
\end{equation}
where $m_p$ is the sky temperature in HEALPix~\cite{Gorski2005} pixel $p$, $g_k$ is a ground pickup amplitude in azimuth bin $k$, and $n_i$ is the noise at sample $i$. The pointing matrices $A_{ip}$ and $G_{ik}$ are sparse, with exactly one non-zero entry per row. In matrix notation, 
\begin{equation}
\mathbf{d} = \tilde{\mathbf{A}} \mathbf{x} + \mathbf{n},
\end{equation}
where $\tilde{\mathbf{A}} = [\mathbf{A} \,|\, \mathbf{G}]$ is a generalized pointing matrix and $\mathbf{x} = (\mathbf{m}, \mathbf{g})^T$. If we assume Gaussian noise with covariance $\mathbf{N}$, the maximum-likelihood solution minimizes the $\chi^2$ statistic, yielding the normal equations
\begin{equation}
	\tilde{\mathbf{A}}^T \mathbf{N}^{-1} \tilde{\mathbf{A}} \, \mathbf{x} = \tilde{\mathbf{A}}^T \mathbf{N}^{-1} \mathbf{d}.
	\label{eq:normal_eq}
\end{equation}
\begin{figure}[b!]
	\centering
	\includegraphics[width=0.8\textwidth]{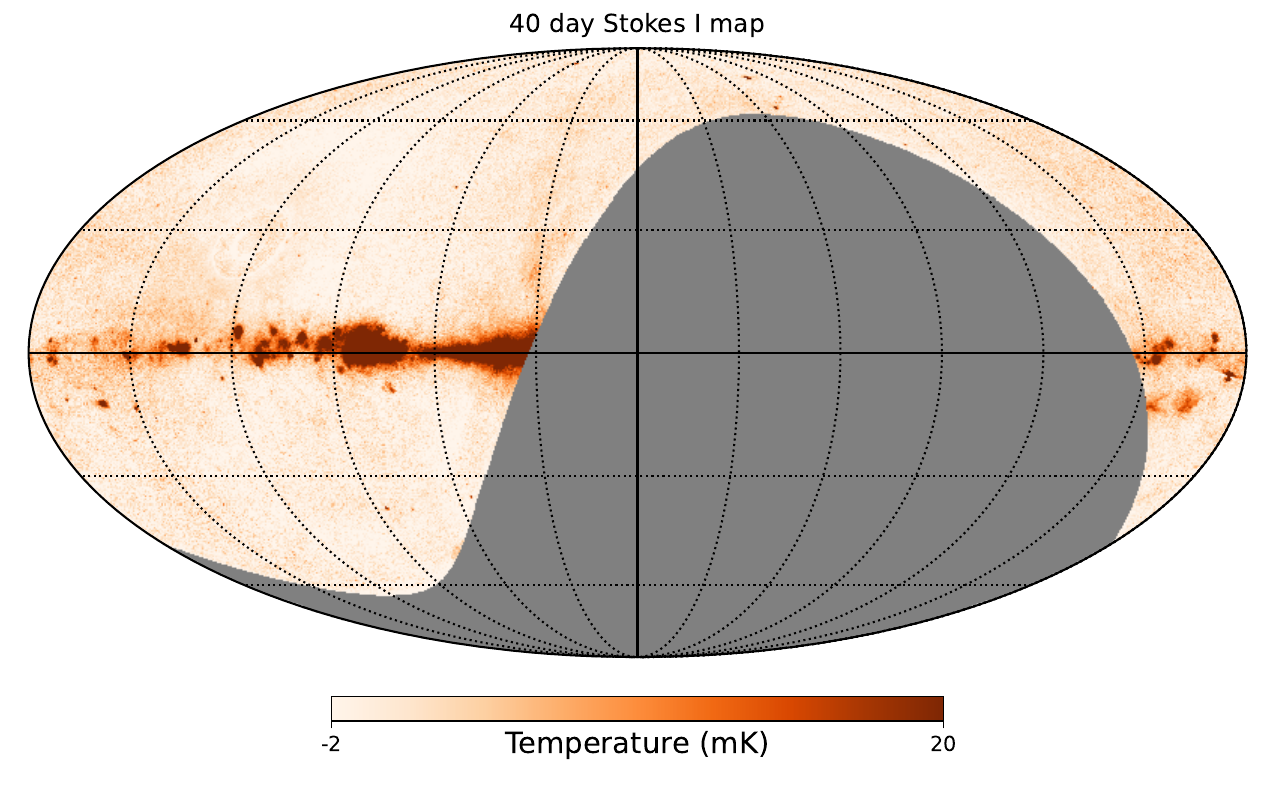}
	\caption{\textbf{Stokes $I$ map of the northern sky at 8--10\,GHz} produced by the CGEM map-making pipeline, shown in Mollweide projection. This map was obtained with $\sim900$ hours of integration time with the warm radiometer. Note the North Polar Spur. The grey region is the presently-unobserved southern sky (declination $<$ 0).}
	\label{fig:stokesI}
\end{figure}
Since the system is too large to invert directly ($N_\text{samples} \sim 10^7$ for a one-day map), we solve it iteratively using a preconditioned conjugate gradient (PCG) algorithm~\cite{Wendland_2017}. This approach is usually found in the literature as generalized least squares (GLS). The noise inverse $\mathbf{N}^{-1}$ is applied efficiently in the Fourier domain, exploiting the Toeplitz structure of the noise covariance for stationary noise segments. We implement an overlap-add filtering method that enables efficient filtering of long-timescale drifts by exploiting the fact that
\begin{equation}
    [\textbf{N}^{-1} \textbf{x}]_{\mathrm{freq}} = \mathcal{F}^{-1} \left( \frac{\mathcal{F}[\textbf{x}](f)}{P(f)} \right).
\end{equation}
A challenge for CGEM's constant-elevation scan strategy is the degeneracy between the average sky signal at declination $\delta_p$ and the ground pickup at azimuth $g_k$: for pixels observed at a single azimuth, the data cannot distinguish between sky emission and ground contamination. To manage this degeneracy we have developed a Tikhonov regularization scheme~\cite{tikhonov1977solutions}, in which the normal equations are replaced by
\begin{equation}
	\left(\tilde{\mathbf{A}}^T \mathbf{N}^{-1} \tilde{\mathbf{A}} + \alpha_{\mathrm sky}^2\,\mathbf{L}_{\mathrm sky}^T\mathbf{L}_{\mathrm sky} + \alpha_{\mathrm gnd}^2\,\mathbf{L}_{\mathrm gnd}^T\mathbf{L}_{\mathrm gnd}\right)\mathbf{x} = \tilde{\mathbf{A}}^T \mathbf{N}^{-1} \mathbf{d},
	\label{eq:tikhonov}
\end{equation}
where $\mathbf{L}_{\mathrm sky}$ and $\mathbf{L}_{\mathrm gnd}$ penalize the the sky signal averaged over right ascension, and the variance of the ground pickup independently, and $\alpha_{\mathrm sky}$, $\alpha_{\mathrm gnd}$ are their respective regularization strengths. A Galactic mask is applied to the sky penalty to avoid biasing bright emission regions. 

\begin{figure}[t!]
	\centering
	\includegraphics[width=\textwidth]{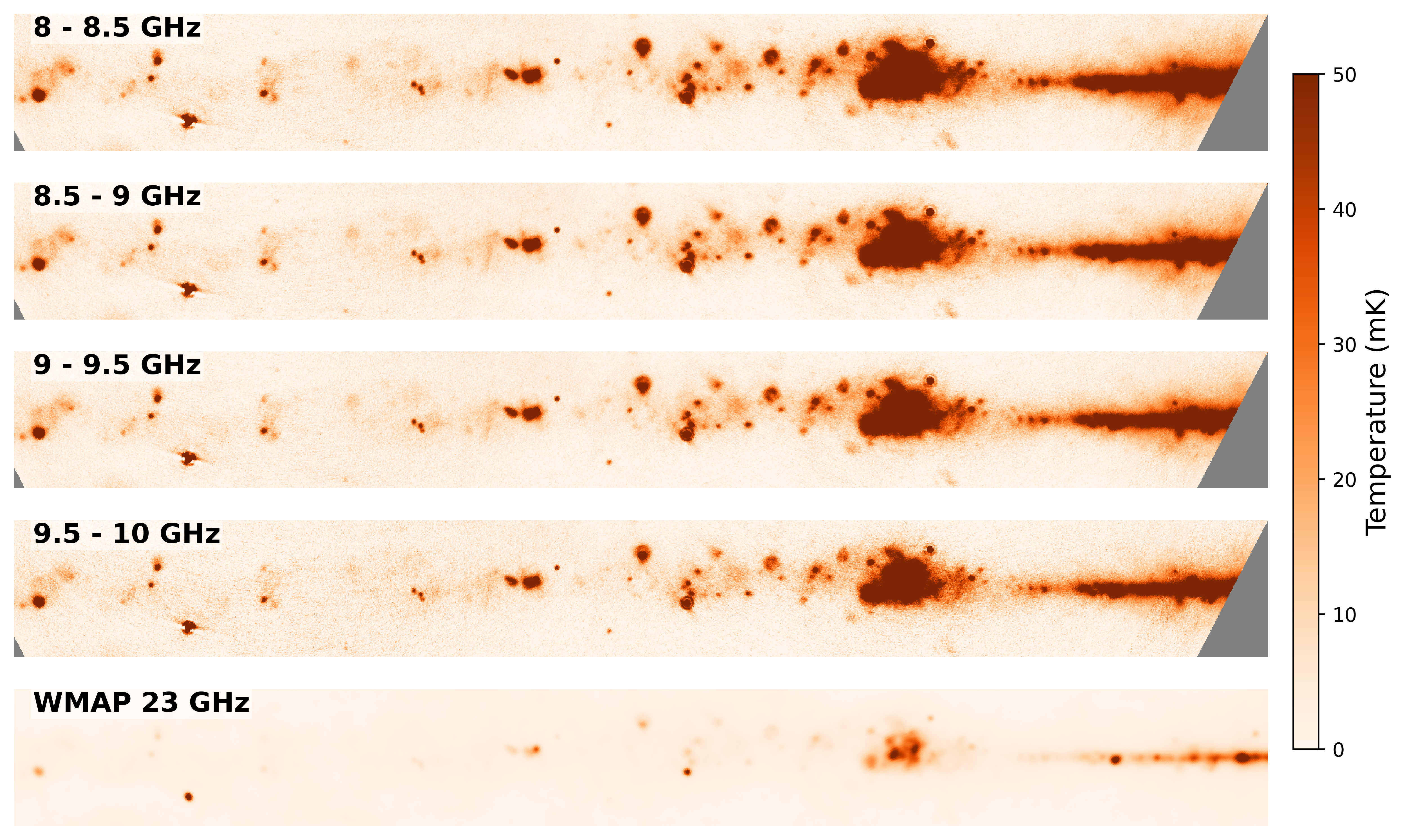}
	\caption{\textbf{Subband maps of the Galactic plane ($27^\circ<\ell<210^\circ$ and $|b|<10^\circ$).} The sub-bands are split into four 500 MHz sub-bands and are displayed in ascendant order and the bottom one is WMAP 23 GHz map from~\citenum{Bennett_2013}. We can already see frequency scaling of the emission in the plane (which is brighter at lower frequencies). 
	}
	\label{fig:subbands}
\end{figure}

The first results produced with this pipeline are shown in Figures~\ref{fig:stokesI} (full band-average) and~\ref{fig:subbands} (four subbands). These maps are based on $\sim900$ hours of data with a warm radiometer and produced our first northern sky Stokes I maps. These maps are being actively analyzed and are helping us refine our calibration system. They present a promising step towards the polarization maps that are the primary goal of CGEM.

\section{conclusion}
We have designed and built a custom 4-m radio telescope called CGEM that operates in the 8-10 GHz band with $\sim0.5$ degree angular resolution. CGEM is optimized to measure polarized emission from our galaxy with a specific goal of aiding in the search for B-mode polarization in the CMB. Currently CGEM is being commissioned with a prototype, room-temperature radiometer, but we are simultaneously developing a cryogenic receiver to achieve our target sensitivity. From its present site near Penticton BC, CGEM will map the northern sky (dec $>-5$ degrees). Once the cryogenic system is commissioned, we propose to build a copy of CGEM to deploy in the southern hemisphere to cover the regions being mapped by the most sensitive CMB experiments.

In this paper we have described the overall design of CGEM, with an emphasis on the radiometer design and performance. In a companion paper, we describe the design and performance of the CGEM optics. We also present here the first maps of unpolarized emission from $\sim$900 hours of data taken with the prototype warm radiometer. These early results give us confidence that CGEM will ultimately meet its goal of providing valuable foreground templates for forthcoming CMB B-mode searches.


\acknowledgments 
CGEM is located on the traditional, ancestral and unceded territory of the syilx  people, within the grounds of the Dominion Radio Astrophysical Observatory (DRAO).  We benefit enormously from the stewardship of the land by the syilx Okanagan Nation and from the radio interference protection work of the syilx Okanagan Nation and the National Research Council (NRC). We thank the DRAO for their gracious hospitality and their expertise. CGEM was funded by a grant from the Canada Foundation for Innovation (CFI) 2018 Leading Edge Fund (Project 36483) and  by contributions from
the province of British Columbia and the Natural Science and Engineering Research Council (NSERC) through Discovery Grants. This research was enabled in part by support provided by
CalculQuebec \href{(https://www.calculquebec.ca/en/)}{https://www.calculquebec.ca/en/} and the Digital Research Alliance of Canada
\href{(alliancecan.ca)}{alliancecan.ca}. We gratefully acknowledge the Collaboration for Astronomy Signal Processing and Electronics Research (CASPER) for providing open-source hardware and software development tools
that were essential to the signal processing infrastructure used in this research. PVG received the support of a fellowship from “la Caixa” Foundation (ID 100010434,  code  B005800), a Rafel del Pino Excellence Scholarship and a UBC 4YF Doctoral Fellowship. PZ received the support of a UBC 4YF Doctoral Fellowship. J.M. acknowledges the support of the Natural Sciences and Engineering Research Council of Canada (NSERC)
[funding reference number 569654]. CGEM has made use of the following Open Source Software:~\citenum{pandas_paper, OpenMPI, healpy, Grafana, Prometheus, matplotlib_zenodo, SciPy, NumPy, ducc, bitshuffle}.


\bibliography{report} 
\bibliographystyle{spiebib} 

\end{document}